\documentclass[a4paper,11pt]{article}
\usepackage{pos}

\usepackage[british]{babel}
\usepackage{xspace}
\usepackage[output-decimal-marker={,}, binary-units = true]{siunitx}
\usepackage{chngcntr}
\usepackage{cleveref}
\crefname{table}{table}{tables}
\Crefname{table}{Table}{Tables}
\crefname{figure}{figure}{figures}
\Crefname{figure}{Fig.}{Figures}

\makeatletter

\newcommand{\ET}{\ensuremath{E_{\text{T}}}\xspace}
\newcommand{\ETmiss}{\ensuremath{\ET\hspace{-1.1em}/\kern0.45em}\xspace}

\makeatother

\newlength{\bibitemsep}
\newlength{\bibparskip}
\let\oldthebibliography\thebibliography
\renewcommand\thebibliography[1]{%
  \oldthebibliography{#1}%
  \setlength{\parskip}{\bibitemsep}%
  \setlength{\itemsep}{\bibparskip}%
}

\title{Exploring jets: substructure and flavour tagging in CMS and ATLAS}
\ShortTitle{Jets at the LHC}

\author*[a]{Andrea Malara}

\affiliation[a]{Universite Libre de Bruxelles (ULB),\\
Av. Franklin Roosevelt 50, 1050 Bruxelles, Belgium}

\emailAdd{andrea.malara@cern.ch}

\abstract{
The identification and characterization of jets are crucial tasks for effectively probing
fundamental particle interactions. The ATLAS and CMS experiments have developed cutting-edge techniques
to improve jet identification and calibration, employing innovative approaches including
advanced neural network architectures, attention-based mechanisms, and adversarial training.
These proceedings provide a comprehensive review of the state-of-the-art methods employed by both collaborations,
highlighting their similarities, unique strengths, and limitations through a comparative analysis.
}

\FullConference{12th Large Hadron Collider Physics Conference  (LHCP2024)\\
3-7 June 2024 \\
Boston, USA \\}

\begin{document}
\maketitle

\section{Introduction}

At the Large Hadron Collider (LHC), proton-proton (pp) collisions produce numerous quarks and gluons,
which almost immediately undergo fragmentation and hadronization.
This process results in collimated showers of hadrons, commonly referred to as jets,
which are detected as clusters of energy deposits in the detectors.
Accurate jet reconstruction and calibration are crucial to ensure the success of the LHC physics program.

Understanding the origins of jets, particularly distinguishing between those initiated by quarks or gluons,
is crucial for accurate identification.
Additionally, the hadronic decay products of highly energetic, heavy particles,
like the top quark or massive bosons, can be reconstructed as a single large-radius jet
when the originating particle is strongly Lorentz-boosted.
Such jets are characterized by a distinctive radiation pattern, often referred to as jet substructure,
and specific flavour content, which can be used to discriminate between different underlying physics processes.

These proceedings will present and compare the state-of-the-art machine learning (ML) techniques employed
by the CMS~\cite{CMS:2008xjf} and ATLAS~\cite{ATLAS:2008xda} experiments for jet calibration and identification.
A detailed comparison will be provided, highlighting the major differences,
common tools, and approaches used by both collaborations.

\section{Jet tagging}

Jet identification (or tagging) aims at distinguishing the type of particle
and its subsequent decay that produced a jet by analyzing the jet's substructure and flavour composition.

\begin{figure}[t]
\centering
\includegraphics[width=0.38\textwidth]{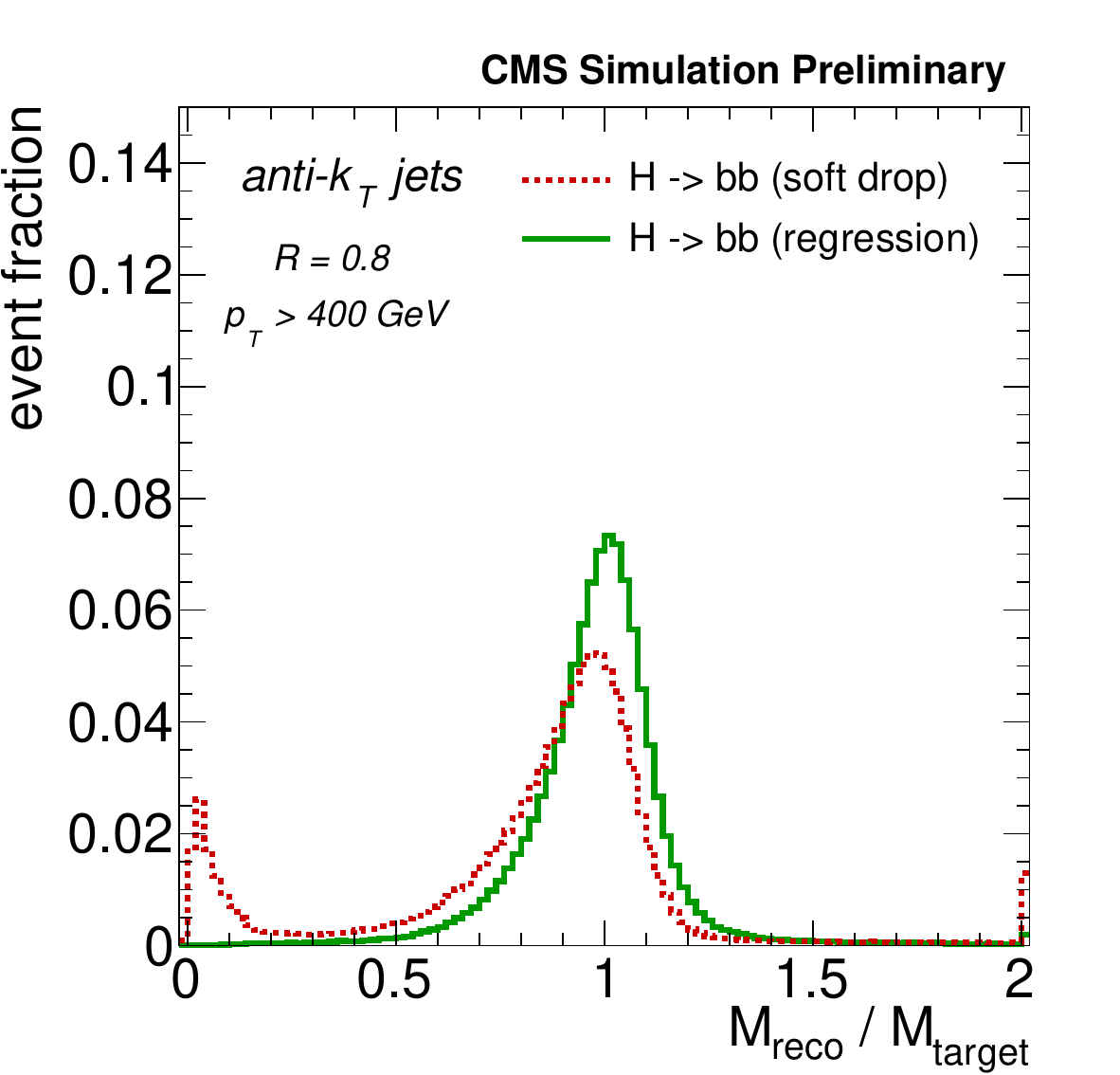}
\includegraphics[width=0.46\textwidth]{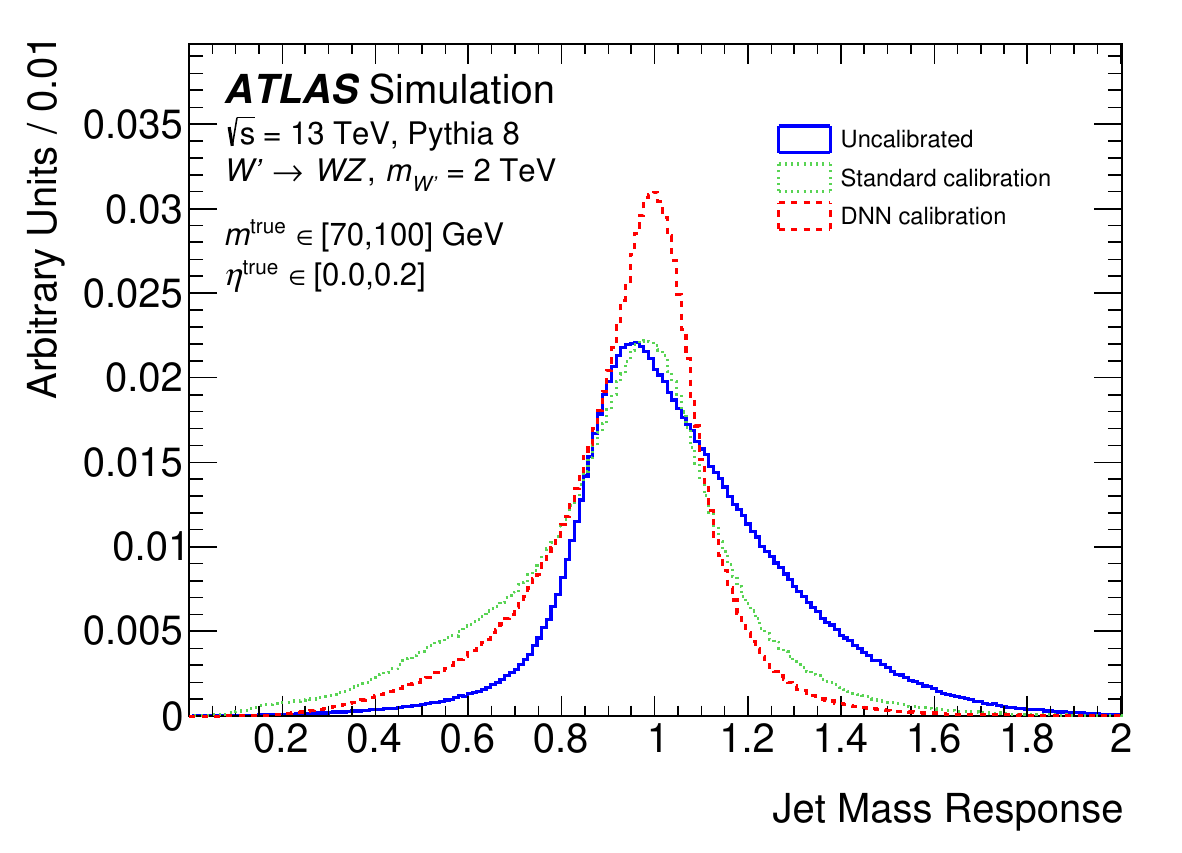}
\caption{Jet mass response for jets originating from massive bosons using regression algorithms by CMS (left) and ATLAS (right) collaborations, respectively.
Both mass regression algorithms show a substantial improvement in the mass resolution and in the absolute scale. Taken from~\cite{CMS-DP-2021-017,atlascollaboration2023simultaneousenergymasscalibration}.
}
\label{fig:MassRegression}
\end{figure}

The jet mass is one of the most important observables for identifying jets originating from the decays of heavy particles, such as W, Z, and Higgs bosons, as well as the top quark.
While the peak position of the jet mass is often accurately reconstructed,
the mass resolution is typically large, of the order of 10\%, mostly limited by the detector resolution.
Advanced calibration techniques based on ML have been developed, resulting in an improvement in mass resolution by 10-20\%,
as shown in \cref{fig:MassRegression}.
Both CMS and ATLAS collaborations adopted similar techniques for large-radius jets~\cite{CMS-DP-2021-017,atlascollaboration2023simultaneousenergymasscalibration}.
The technique used by the ATLAS Collaboration involves a series of deep neural network (DNN) architectures,
designed to perform a simultaneous regression of both the energy and mass of large-radius jets,
exploiting high-level jet variables, as well as event-level variables.
Conversely, the approach used by the CMS Collaboration is based on ParticleNet~\cite{Qu_2020},
a graph neural network (GNN) based on the Dynamic Graph Convolutional Neural Network approach
that takes an unordered set of jet constituent particles as input, redesigned for both mass regression and flavour tagging,
particularly to distinguish jets originating from massive bosons and those produced by gluon splitting.

\begin{figure}[t]
\centering
\includegraphics[width=0.40\textwidth]{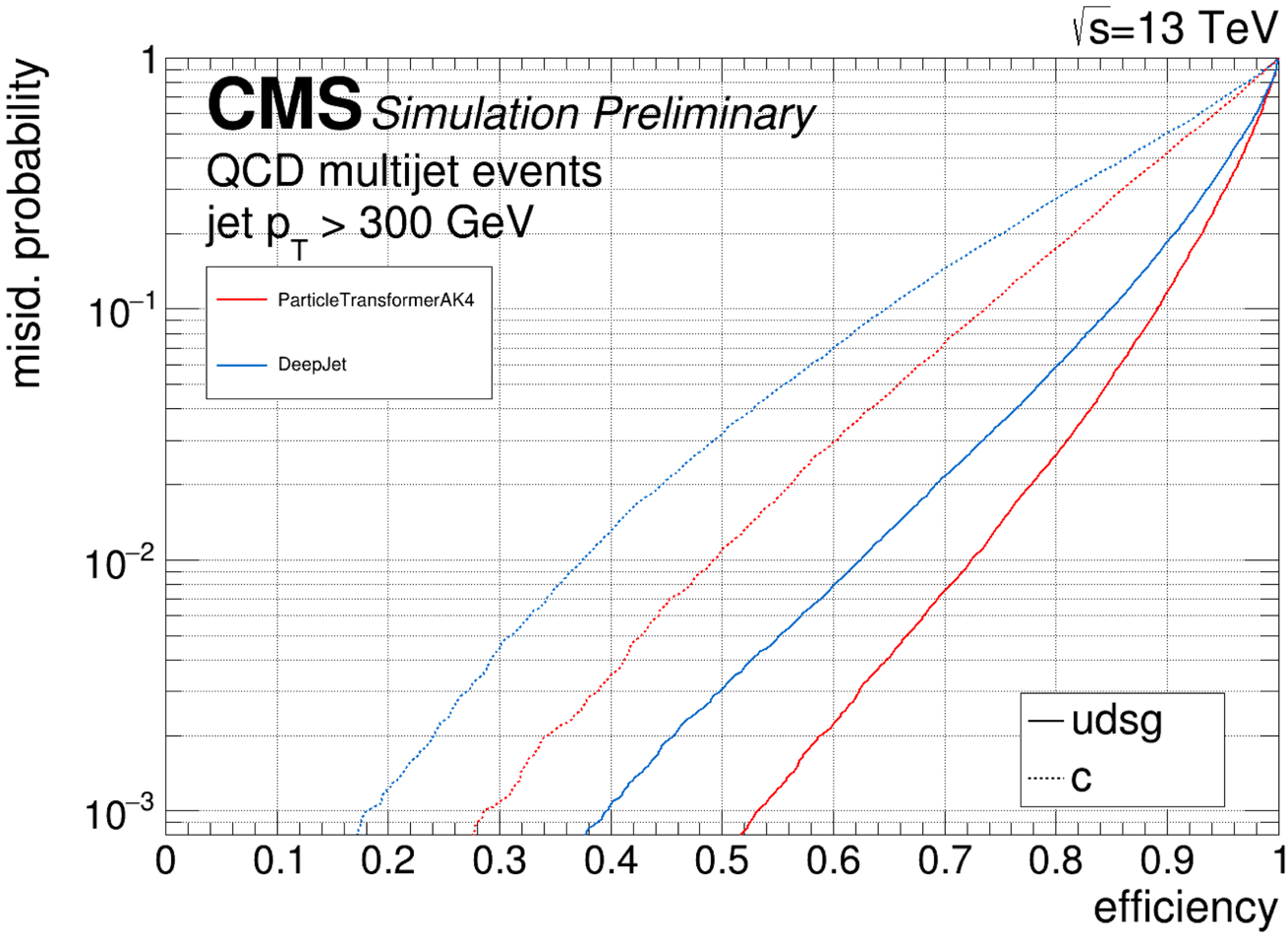}
\includegraphics[width=0.40\textwidth]{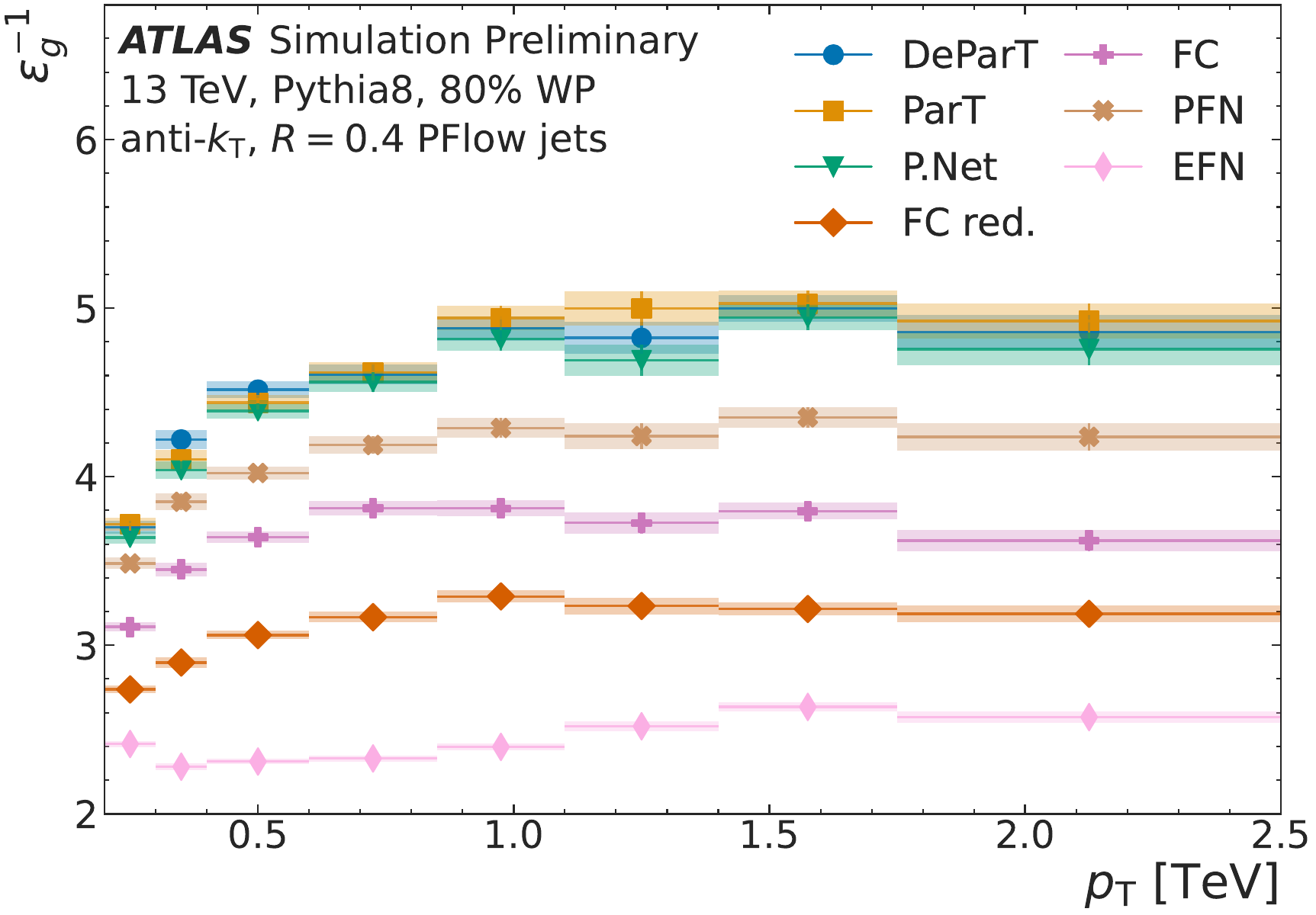}
\caption{
Performance comparison of the Particle Transformer with other algorithms.
Left: Probability of misidentifying non-b-quark initiated jets as a function of the efficiency for identifying b-quark-initiated jets, used by the CMS Collaboration. Taken from ~\cite{CMS-DP-2022-050}
Right: Rejection rate of gluon-initiated jets as a function of jet transverse momentum for a fixed 80\% selection probability of light-quark jets, used by the ATLAS Collaboration. Taken from ~\cite{ATL-PHYS-PUB-2023-032}.
}
\label{fig:PartT}
\end{figure}

One of the latest advancements in ML architectures is the use of attention-based (transformer) mechanisms,
which have been swiftly adopted by both the ATLAS and CMS collaborations.
Such architecture, referred to as Particle Transformer~\cite{qu2024particletransformerjettagging} by both experiments,
represents a paradigm shift in jet tagging compared to traditional DNN approaches, as it better captures
both local and global relationships between particles within a jet.
This newer architecture has demonstrated improved performance in jet tagging tasks,
particularly in environments with many overlapping particles, and is effectively used
to distinguish jets originating from gluons, light-flavour quarks, and heavy-flavour quarks in both experiments~\cite{CMS-DP-2022-050,ATL-PHYS-PUB-2023-032}. Examples of the improved performances of Particle Transformer compared to various algorithms are reported in \cref{fig:PartT}.

\begin{figure}[t]
\centering
\includegraphics[width=0.38\textwidth]{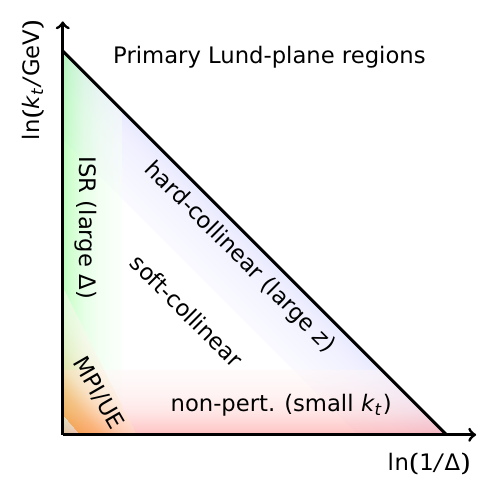}
\includegraphics[width=0.38\textwidth]{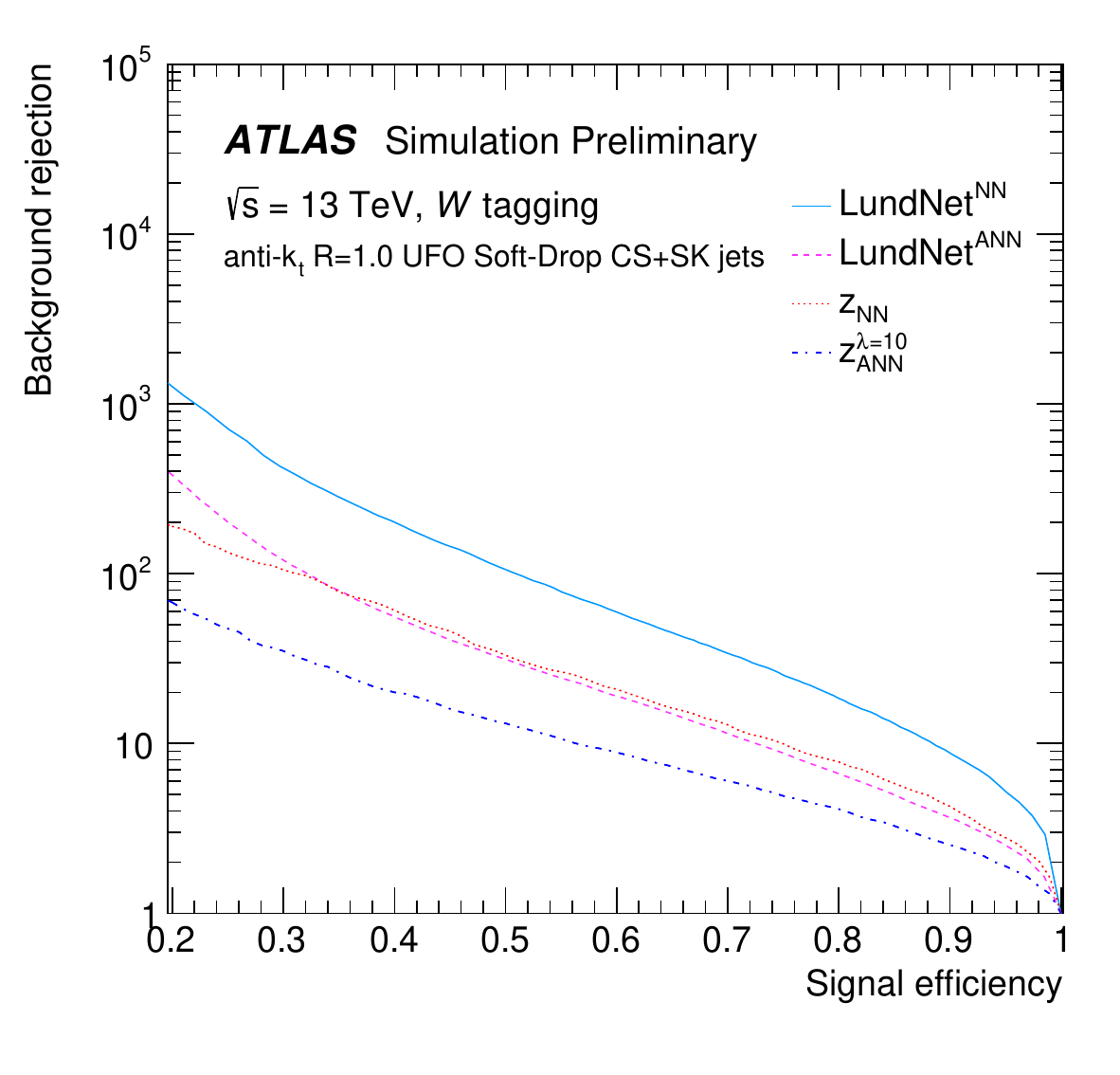}
\caption{
Left: Schematic representation of the different regions of the Lund jet plane. Taken from ~\cite{Dreyer_2018}.
Right: Rejection rate of QCD-initiated jets as a function of the selection efficiency for W-boson-initiated jets.
High-level-feature-based taggers, denoted with $z_{NN}$, are shown as reference. Taken from ~\cite{ATL-PHYS-PUB-2023-017}.
}
\label{fig:LundPlane}
\end{figure}

A diverse approach to jet tagging uses a theoretically well-motivated representation of the jet's radiation pattern
during the parton shower process, provided by the Lund jet plane~\cite{Dreyer_2018}.
As illustrated in \cref{fig:LundPlane} (left), this representation is defined by two key variables:
the transverse momentum of emitted particles and the angular separation between emissions.
LundNet~\cite{Dreyer_2018}, a machine learning-based jet tagging algorithm deployed by the ATLAS experiment,
leverages the Lund jet plane to capture the sequential splitting of particles within a jet,
such as the production of secondary quarks or gluons, revealing characteristic patterns
of initiating particle (e.g., quark, gluon, or heavy particle decay).
The ability of LundNet to model the hierarchical structure of jet emissions
enhances tagging performance compared to high-level-feature-based taggers, as demonstrated in \cref{fig:LundPlane} (right).

\section{Calibration techniques}

Recent advancements in ML have led to the development of increasingly sophisticated tools and algorithms for jet tagging.
These state-of-the-art techniques allow us to probe deeper into jet properties, such as substructure,
and identify new features that were so far hidden or not well understood.
This exploration paves the way into uncharted territories, with the potential to reveal new phenomena in particle physics.
However, as we push the boundaries of such remarkable tools, it becomes essential to assess
their sensitivity to the specific details of simulated data and to understand their reliability and limitations
in accurately describing pp collision data.

\begin{figure}[t]
    \centering
\includegraphics[width=0.45\textwidth]{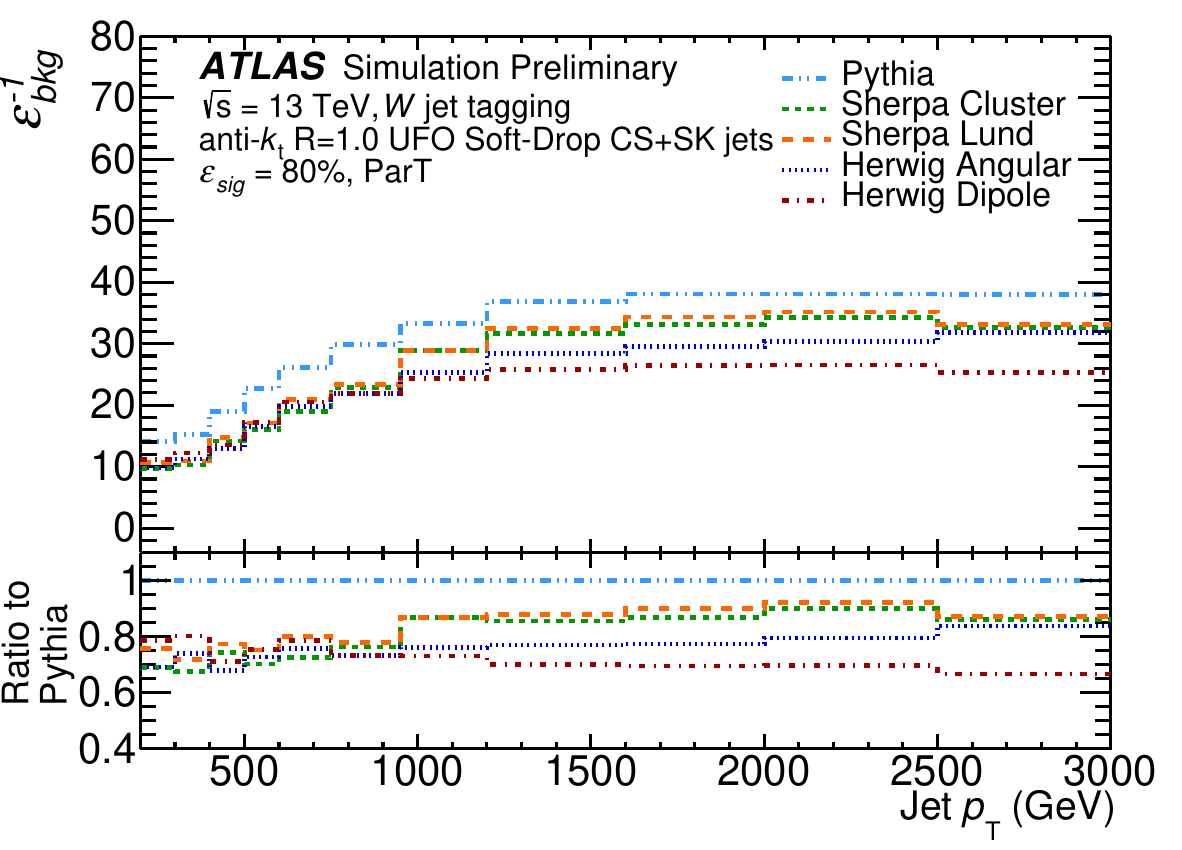}
 \includegraphics[width=0.38\textwidth]{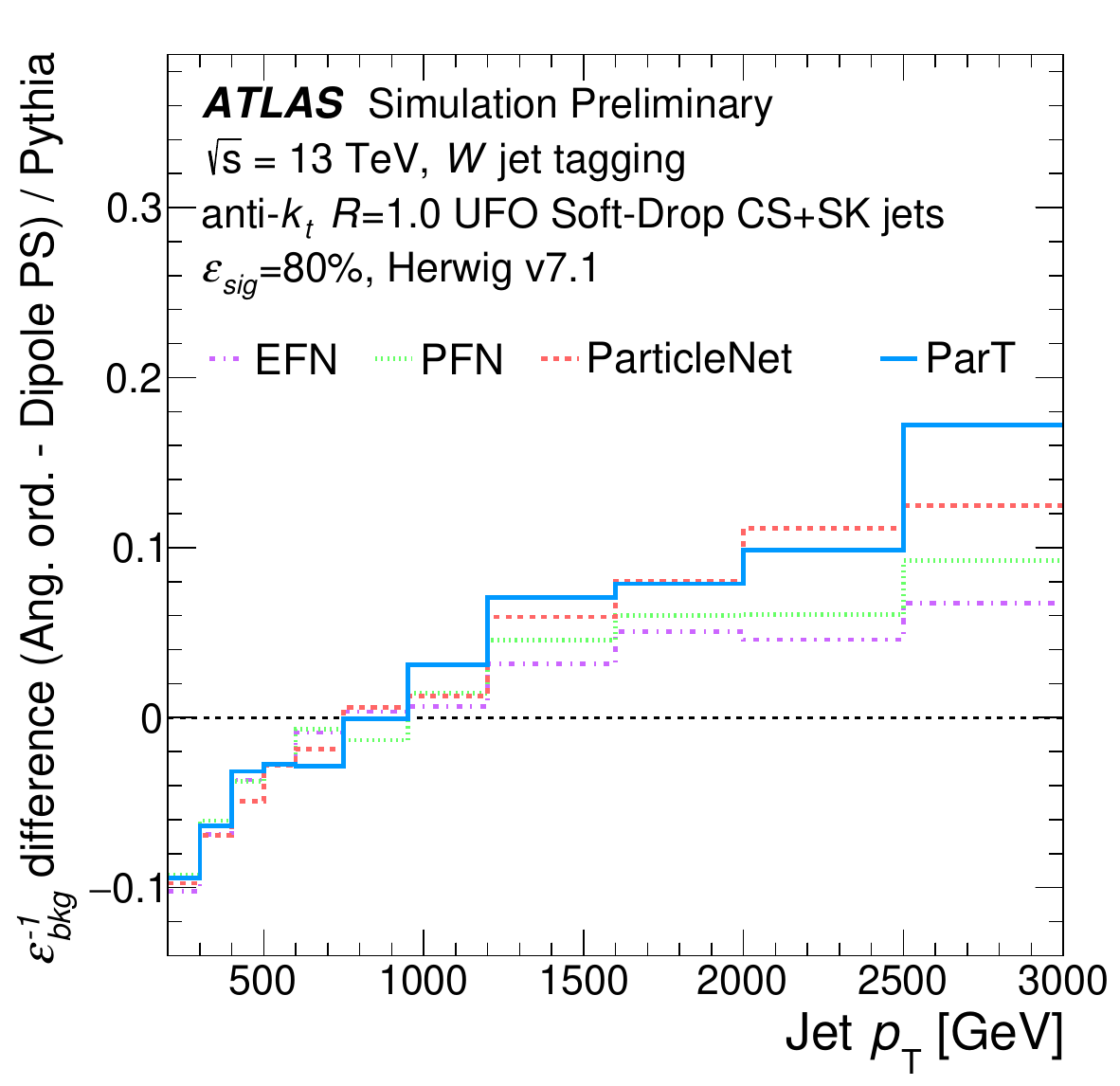}
\caption{
Left: Rejection rate of W-boson-initiated jets as a function of jet transverse momentum for different simulated samples using the Particle Transformer algorithm in ATLAS.
Right: Difference in rejection rate as a function of jet transverse momentum for various jet taggers, illustrating sensitivity to hadronization and parton shower modeling. Taken from ~\cite{ATL-PHYS-PUB-2023-020}.
}
    \label{fig:ModellingSimulation}
\end{figure}

The ATLAS Collaboration has recently conducted a series of studies~\cite{ATL-PHYS-PUB-2023-017,ATL-PHYS-PUB-2023-020} 
to assess the performance of these algorithms across different simulation scenarios (see \cref{fig:ModellingSimulation} (left))
and to evaluate their robustness against mismodeling effects related to hadronization and parton shower (see \cref{fig:ModellingSimulation} (right)).
These studies highlight the importance of rigorous validation of ML models
to ensure their reliability when applied to real collision data.

Several advanced techniques have been developed to improve the modeling of various jet taggers
and ensure consistency with collision data.
One notable method involves adversarial training, which enhances the robustness of ML models by reducing
their sensitivity to systematic uncertainties and improving their generalization across different simulations.
Used in both ATLAS~\cite{ATL-PHYS-PUB-2023-020} and CMS~\cite{CMS-DP-2022-049} experiments, this approach increases the reliability of jet tagging algorithms,
resulting in improved performance when applying models trained on simulated data to real pp events.
While adversarial training directly improves the output of ML taggers, residual discrepancies between data and simulations
can be corrected a posteriori using reweighting techniques. These corrections are crucial for ensuring accurate comparisons
of the performance of different jet tagging algorithms using real pp collision data.

\begin{figure}[t]
\centering
\includegraphics[width=0.40\textwidth]{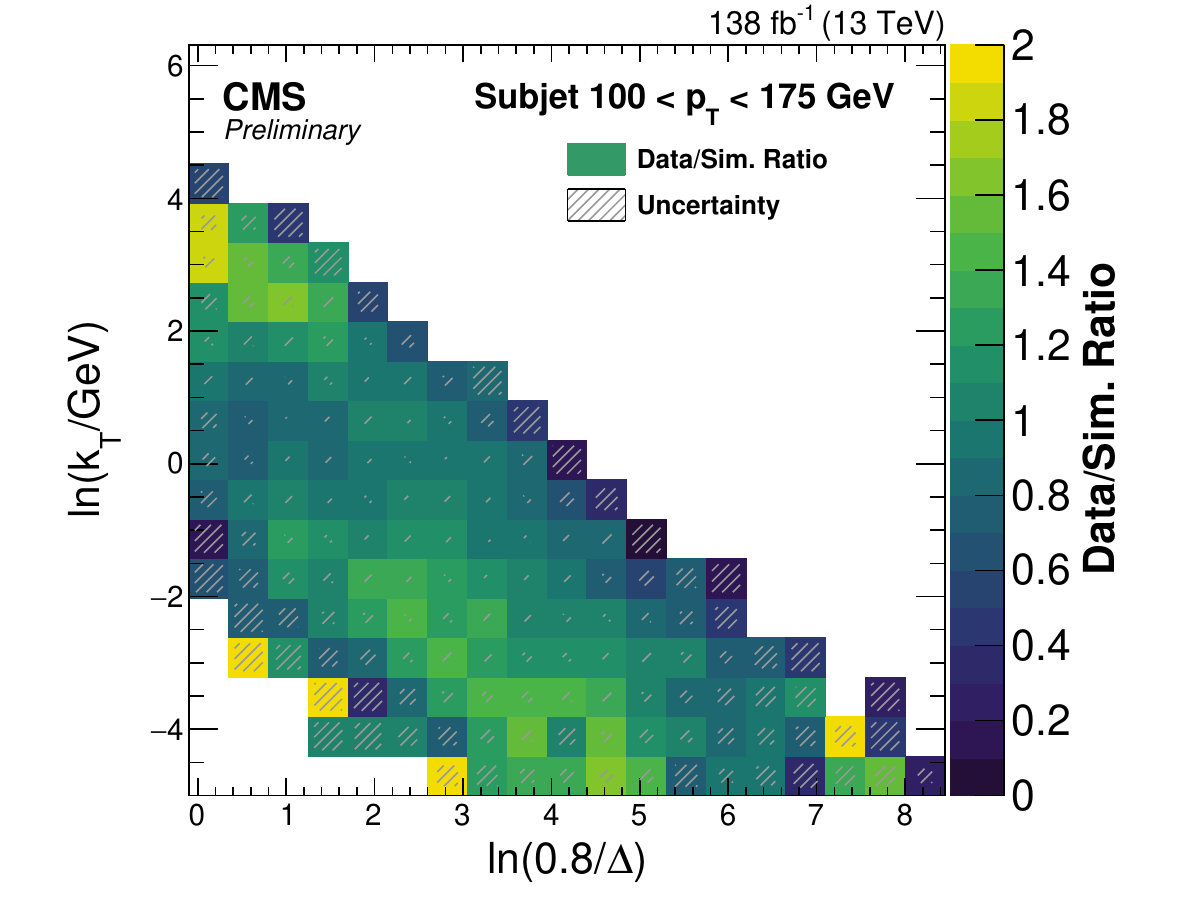}
\includegraphics[width=0.40\textwidth]{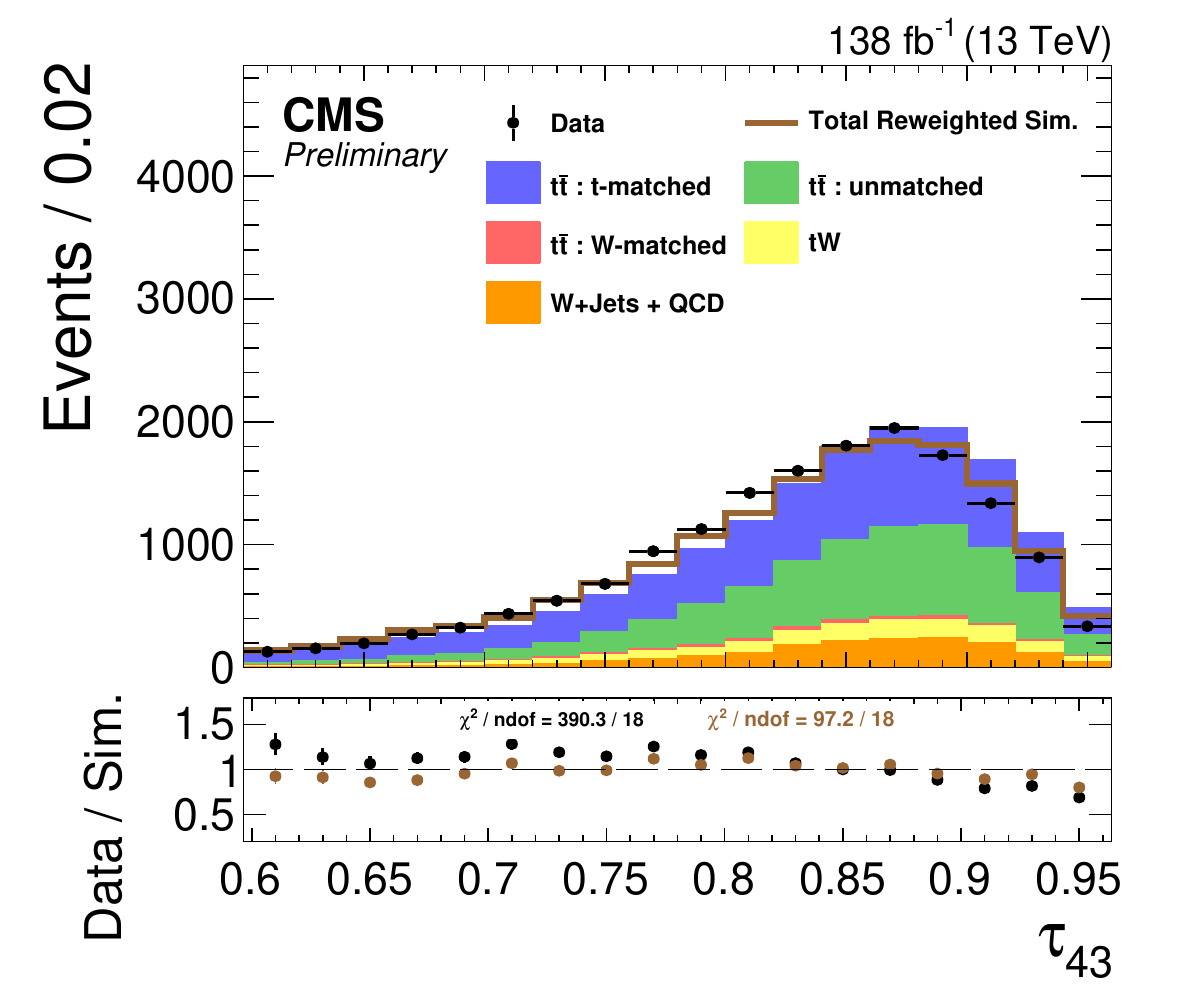}
\caption{
Left: Ratios of the Lund jet plane in data and simulation in a given range of subjet transverse momentum.
Right: Data-to-simulation agreement of a substructure observable. The brown line shows the total simulated distribution after the Lund jet plane correction has been applied. The agreement between data and simulation improves with the correction. Taken from ~\cite{CMS-DP-2023-046}.
}
\label{fig:CMS_LPR}
\end{figure}

An innovative technique developed by the CMS Collaboration involves using the Lund jet plane
to iteratively correct data-to-simulation discrepancies at each stage of jet evolution (see \cref{fig:CMS_LPR} (left)).
This method offers a more granular approach for correcting jet substructure mismodelling,
providing improved precision compared to more traditional reweighting methods.
The technique is versatile and applied to any jet substructure, including rare multi-pronged jets,
and is particularly effective in improving the modeling of soft emissions and non-perturbative effects,
areas where theoretical predictions often show discrepancies with experimental data (see \cref{fig:CMS_LPR} (right)).

\section{Summary}

Accurate jet reconstruction and identification are crucial for the LHC physics program, requiring
advanced machine learning techniques to distinguish jets originating from quarks, gluons, or heavy particles.

These proceedings provide an overview of the state-of-the-art techniques employed by the ATLAS and CMS collaborations
for jet calibration and tagging. Both experiments have adopted similar strategies for jet energy and mass regression
and flavour tagging, utilizing advanced neural network architectures, attention-based mechanisms, and
adversarial training, to improve jet tagging performance. Both collaborations have conducted extensive studies
to ensure reliable performance of the various algorithms when applied to real collision data.

Additionally, innovative approaches based on the Lund jet plane representation have been adopted by both collaborations.
The ATLAS Collaboration uses the LundNet algorithm to improve jet tagging, while the CMS Collaboration employs the Lund jet plane to improve
the modeling of substructure variables.

The similarities and complementarities of the approaches adopted by ATLAS and CMS are summarized,
providing a comprehensive view of the current state-of-the-art techniques for jet calibration and tagging at the LHC.

\bibliographystyle{JHEP}
\bibliography{bibliography}

\providecommand{\href}[2]{#2}\begingroup\raggedright\begin{thebibliography}{10}

\bibitem{CMS:2008xjf}
{\scshape CMS} collaboration, \emph{The {CMS} experiment at the {CERN} {LHC}},
  \href{https://doi.org/10.1088/1748-0221/3/08/S08004}{\emph{JINST} {\bfseries
  3} (2008) S08004}.

\bibitem{ATLAS:2008xda}
{\scshape ATLAS} collaboration, \emph{{The ATLAS Experiment at the CERN Large
  Hadron Collider}},
  \href{https://doi.org/10.1088/1748-0221/3/08/S08003}{\emph{JINST} {\bfseries
  3} (2008) S08003}.

\bibitem{CMS-DP-2021-017}
{\scshape CMS} collaboration, \emph{{Mass regression of highly-boosted jets
  using graph neural networks}},
  \href{https://arxiv.org/abs/CMS-DP-2021-017}{{\ttfamily CMS-DP-2021-017}},
  {https://cds.cern.ch/record/2777006}.

\bibitem{atlascollaboration2023simultaneousenergymasscalibration}
{\scshape ATLAS} collaboration, \emph{Simultaneous energy and mass calibration
  of large-radius jets with the atlas detector using a deep neural network},
  \href{https://arxiv.org/abs/2311.08885}{{\ttfamily 2311.08885}},
  https://arxiv.org/abs/2311.08885.

\bibitem{Qu_2020}
H.~Qu and L.~Gouskos, \emph{Jet tagging via particle clouds},
  \href{https://doi.org/10.1103/PhysRevD.101.056019}{\emph{Phys. Rev. D}
  {\bfseries 101} (2020) 056019}
  https://link.aps.org/doi/10.1103/PhysRevD.101.056019.

\bibitem{qu2024particletransformerjettagging}
H.~Qu, C.~Li and S.~Qian, \emph{Particle transformer for jet tagging},
  {\emph{{Proceedings of the 39th International Conference on Machine Learning,
  PMLR 162:18281-18292, 2022}} (2024) }
  [\href{https://arxiv.org/abs/2202.03772}{{\ttfamily 2202.03772}}],
  https://arxiv.org/abs/2202.03772.

\bibitem{CMS-DP-2022-050}
{\scshape CMS} collaboration, \emph{{Transformer models for heavy flavor jet
  identification}},  \href{https://arxiv.org/abs/CMS-DP-2022-050}{{\ttfamily
  CMS-DP-2022-050}}, https://cds.cern.ch/record/2839920.

\bibitem{ATL-PHYS-PUB-2023-032}
{\scshape ATLAS} collaboration, \emph{{Constituent-Based Quark Gluon Tagging
  using Transformers with the ATLAS detector}},
  \href{https://arxiv.org/abs/ATL-PHYS-PUB-2023-032}{{\ttfamily
  ATL-PHYS-PUB-2023-032}}, https://cds.cern.ch/record/2878932.

\bibitem{Dreyer_2018}
F.A.~Dreyer, G.P.~Salam and G.~Soyez, \emph{The lund jet plane},
  \href{https://doi.org/10.1007/jhep12(2018)064}{\emph{Journal of High Energy
  Physics} {\bfseries 2018} (2018) 64} https://doi.org/10.1007/JHEP12(2018)064.

\bibitem{ATL-PHYS-PUB-2023-017}
{\scshape ATLAS} collaboration, \emph{Tagging boosted $w$ bosons applying
  machine learning to the lund jet plane},
  \href{https://arxiv.org/abs/ATL-PHYS-PUB-2023-017}{{\ttfamily
  ATL-PHYS-PUB-2023-017}}, https://cds.cern.ch/record/2864131.

\bibitem{ATL-PHYS-PUB-2023-020}
{\scshape ATLAS} collaboration, \emph{{Constituent-Based $W$-boson Tagging with
  the ATLAS Detector}},
  \href{https://arxiv.org/abs/ATL-PHYS-PUB-2023-020}{{\ttfamily
  ATL-PHYS-PUB-2023-020}}, https://cds.cern.ch/record/2866592.

\bibitem{CMS-DP-2022-049}
{\scshape CMS} collaboration, \emph{{Adversarial training for b-tagging
  algorithms in CMS}},  \href{https://arxiv.org/abs/CMS-DP-2022-049}{{\ttfamily
  CMS-DP-2022-049}}, https://cds.cern.ch/record/2839919.

\bibitem{CMS-DP-2023-046}
{\scshape CMS} collaboration, \emph{{Lund Plane Reweighting for Jet
  Substructure Correction}},
  \href{https://arxiv.org/abs/CMS-DP-2023-046}{{\ttfamily CMS-DP-2023-046}},
  https://cds.cern.ch/record/2866330.

\end{thebibliography}\endgroup

\end{document}